\documentclass{PoS}
\usepackage{amsmath,graphicx,float}

\title{Spectral properties of light and charm mesons
from $N_f=2+1$ anisotropic lattice QCD}

\ShortTitle{Spectral properties of light and charm mesons}

\author{\speaker{Ryan Quinn}\\
        Department of Theoretical Physics, National University of Ireland Maynooth, Maynooth, Co.Kildare, Ireland\\
        E-mail: \email{ryanquinn@thphys.nuim.ie}}

\author{Jonas Glesaaen\\
        Department of Physics, College of Science, Swansea University, Swansea SA28PP,      UK\\
        E-mail: \email{jonas.glesaaen@swansea.ac.uk}}
        
\author{Alexander Rothkopf\\
        Faculty of Science and Technology, University of Stavanger, NO-4036 Stavanger, Norway\\
        E-mail: \email{alexander.rothkopf@uis.no}}
        
\author{Jon-Ivar Skullerud\\
        Department of Theoretical Physics, National University of Ireland Maynooth, Maynooth, Co.Kildare, Ireland\\
        E-mail: \email{jonivar@thphys.nuim.ie}}

\abstract{We compute temporal correlators and spectral functions for light, open
charm and charmonium mesons in the pseudoscalar and vector channel for
a range of temperatures below and above the deconfinement transition.
The study is carried out using anisotropic lattice QCD with 2+1
dynamical flavours, $a_s=0.123\,$fm and $a_s/a_\tau=3.5$.  The high-temperature
results are benchmarked by comparing them to reconstructed correlators
obtained by direct summation of the zero temperature correlator. We
use two Bayesian methods to reconstruct the spectral functions: the
maximum entropy method and the more recent BR method.}

\FullConference{XIII Quark Confinement and the Hadron Spectrum - Confinement2018\\
		31 July - 6 August 2018\\
		Maynooth University, Ireland}

\begin{document}
\providecommand{\ncfg}{\ensuremath{N_{\text{cfg}}}}
\providecommand{\nsrc}{\ensuremath{N_{\text{src}}}}

\section{Introduction}

There has been a long-standing interest in heavy quarks as probes of
the quark--gluon plasma, and in particular on the survival or
suppression of their bound states.  Traditionally, the focus has been
on quarkonium ($J/\psi$ and $\Upsilon$) states, but in recent years
there has been an increasing interest in open heavy flavour,
including D meson flow \cite{Abelev:2013lca} and yields
\cite{Adamczyk:2014uip,Adam:2015sza,Adam:2015jda}.

There have been numerous theoretical studies of quarkonium, using a
range of methods including potential models and lattice calculations (see, e.g., \cite{Ding:2012sp,Aarts:2014cda,Kelly:2018hsi,Ding:2018uhl} and references therein).
Despite these efforts, there is still no consensus on how far into the
deconfined region the $J/\psi$ meson may survive, with estimates for
the ``melting temperature''\footnote{This is not a precisely defined
  concept, and it is arguably more appropriate to describe the
  temperature dependence of the bound states in terms of a gradual
  weakening and broadening of the spectral functions.  However, a
  ``melting temperature'' can serve as a convenient shorthand.}
ranging from $T_c$ to $1.7T_c$, where $T_c$ is the pseudocritical
temperature.  Further, high-precision studies of the charmonium system
are therefore required.

In contrast, theoretical studies of open-charm systems are still in
their infancy, with only a single lattice study of open-charm temporal
correlators and spectral functions so far \cite{Kelly:2018hsi}.  In
this study, we will improve on the results presented in that paper by
significantly enhancing our statistics.

\section{Methods}
In this study we deploy lattices generated by the FASTSUM collaboration \cite{Aarts:2014cda, Aarts:2014nba} to describe the QCD medium in which the open and hidden heavy flavor mesons are immersed. These second generation ensembles feature 2+1 flavors of anisotropic clover fermions and a mean-field improved anisotropic Szymanzik gauge action. With an anisotropy parameter of $a_s/a_\tau=3.5$ and lattice spacing of $a_s=0.123$ fm. The strange quark mass is tuned to its physical value, while the light quarks correspond to a pion mass $m_{\pi}=380$ MeV.  The lattice volumes and temperatures are given in Table~\ref{tab:lattices}.  The action is identical to that used by the Hadron Spectrum Collaboration\cite{Edwards:2008ja}, and the zero temperature ($N_\tau= 128$) configurations were kindly provided by them.

\begin{table}[b]
\centering
\begin{tabular}{|ccccrrr|}\hline
$N_s$&   $N_\tau$ & $T$ (MeV) & $T/T_c$ & $\ncfg$ & $\nsrc (c\bar{c})$
  & $\nsrc (\bar{c}l,l\bar{l})$\\\hline
16 & 128 & \,44 & 0.24 &  500 & 16 & 16 \\
32 & 48 & \,122 & 0.63 &  125 & 16 & 16 \\
24 &  40 & 141  & 0.76 &  500 & 16 & 10 \\
24 &  36 & 156  & 0.84 &  500 & 16 & 12 \\
24 &  32 & 176  & 0.95 & 1000 & 16 & 16 \\
24 &  28 & 201  & 1.09 & 1000 & 16 & 14 \\
24 &  24 & 235  & 1.27 & 1000 & 16 &  \\
24 &  20 & 281  & 1.52 & 1000 & 16 &  \\
24 &  16 & 352  & 1.90 & 1000 & 16 & \\ \hline
\end{tabular}
\caption{Lattice volumes $N_s^3\times N_\tau$, temperatures $T$,
  number of configurations $\ncfg$ and number of sources $\nsrc$ used
  in this work.  The
  pseudocritical temperature $T_c$ was determined from the inflection
  point of the Polyakov loop~\cite{Aarts:2014nba}.}
\label{tab:lattices}
\end{table}

The Euclidean correlators $G(\tau; T)$ we compute in the lattice simulation are related to the spectral function $\rho(\omega,\tau)$ via the integral relation
\begin{equation}{
G(\tau,T)={\int_{0}^{\infty} d{\omega} {\rho}({\omega};{T})K({\omega}, {\tau}; {T})} }\,,\end{equation}
\begin{equation}
K(\tau,\omega;T)=\frac{\cosh{(\omega(\tau-1/2T))}}{\sinh{(\omega/2T)}}\,. \end{equation}

Inverting this integral equation is an ill-posed problem. Here we employ methods of Bayesian inference to obtain the most likely spectral function for the given data.
    The starting point for this reconstruction method is Bayes' theorem
    \begin{equation}{
     P[{\rho}|{G,I}]=\frac{P[{G}|{\rho},I]P[{\rho}|{I}]}{P[G|I]} \quad,
     }
     \end{equation}

\noindent where $P{\lbrack}{G}{\vert}{\rho},I{\rbrack} = {\exp}{\lbrack}{-L}{\rbrack}$ is the likelihood function encoding how the data were created
    \begin{equation}{
    L[\rho]=\frac{1}{2}{\sum_{ij}^{N_{{\text{cfg}}}}}{\left(G_i-G_i^{\rho} \right)}C_{ij}{\left(G_j-G_j^{\rho} \right)}}.
    \end{equation}
    $P{\lbrack}{\rho}{\vert}I(m){\rbrack} = {\exp}{\lbrack}{\alpha}{S}{\lbrack}{\rho},m{\rbrack}{\rbrack} $ is the prior probability which provides the regularization of the likelihood function.
    The hyperparameter $\alpha$ weights the influence of the data and prior information. Prior information enters the functional $S$ in two ways: the form of $S$ itself favours certain spectra, and $S$ depends on the so called default model $m$, which by definition is the correct spectrum in the absence of data.
    Numerical optimization is then carried out on the posterior $P{\lbrack}{\rho}{\vert}{G},{I}{\rbrack}$ to find the most probable spectrum
    
    \begin{equation}{\frac{\delta}{{\delta}{\rho}}P{\lbrack}{\rho}{\vert}{G},{I}{\rbrack}\Big|_{\substack{{\rho}={\rho}^{\text{BR}}}}=0
    }.
    \end{equation}
We will here present results using the recently developed BR method \cite{Burnier:2013nla}. We have also analysed the data using the well established maximum entropy method, but these results are overall less well-determined than those from the BR method and will not be shown here.
The systematic uncertainty of these Bayesian methods may be assesed by studying the reconstructed correlator, these are the correlators defined as: 
\begin{equation}
G_{rec}({\widetilde{\tau}},T;T_r)={\int_{0}^{\infty} d{\omega} {\rho}({\omega};{T_r})K({\omega},{\tau}; {T})},
    \end{equation}
where $T_r$ is a reference temperature where the spectral function can be reliably constructed. The reconstructed correlator can also be computed directly from the underlying correlator $G(\tau,T_r)$
without having to extract any spectral functions. Using
\begin{equation}
    {\frac{\cosh(\omega(\tau-N/2))}{\sinh(\omega N/2)} } = \sum_{n=0}^{m-1} \frac{\cosh(\omega(\tau+nN+mN/2))}{\sinh(\omega mN/2)} \end{equation}
with 
\begin{equation}T=\frac{1}{a_{\tau}N},\quad T_r=\frac{1}{a_{\tau}N_r},\quad \widetilde{\tau}=\frac{\tau}{a_{\tau}}, \quad \frac{N_r}{N}=m \in \mathbb{Z}  \end{equation}
we find that
\begin{equation}G_{rec}({\widetilde{\tau}},T;T_r)={\sum_{n=0}^{m-1}}G({\widetilde{\tau}}+nN_{\tau},T_r)\label{Grec-sum}\end{equation}
We have used the lowest temperature, corresponding to $N_\tau=128$, as our reference temperature.  For temperatures where $N_\tau$ is not a factor of 128, we have either removed the middle points or padded with zeros; eg for $N_\tau=40$ we have removed the 8 middle points from the $N_\tau=128$ correlator to obtain the reconstructed correlator, while for $N_\tau=36$ we have padded it with zeros in the middle to obtain $N_r=144$.  Since the correlator is exponentially suppressed in the middle of the lattice, this has little practical impact on the results we will show here.

\section{Results}
\begin{figure}[t] 
\includegraphics*[width=0.5\textwidth]{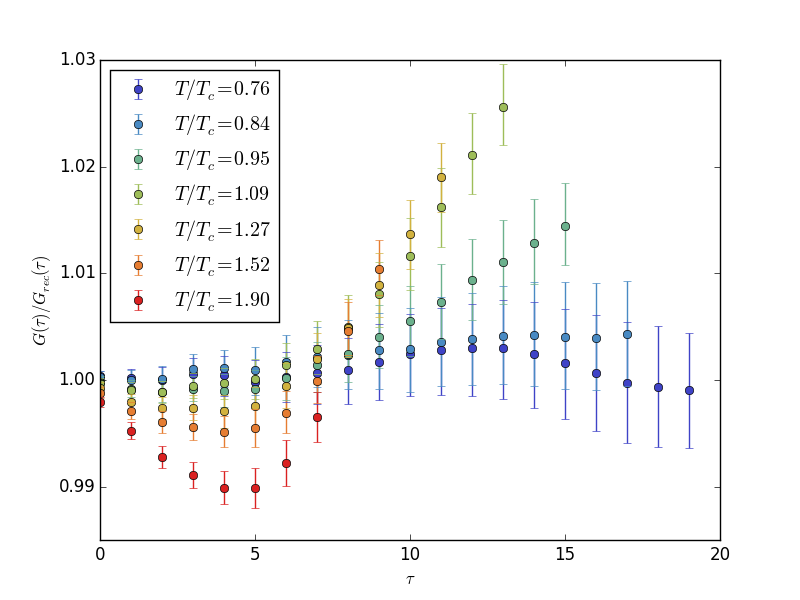}
\includegraphics*[width=0.5\textwidth]{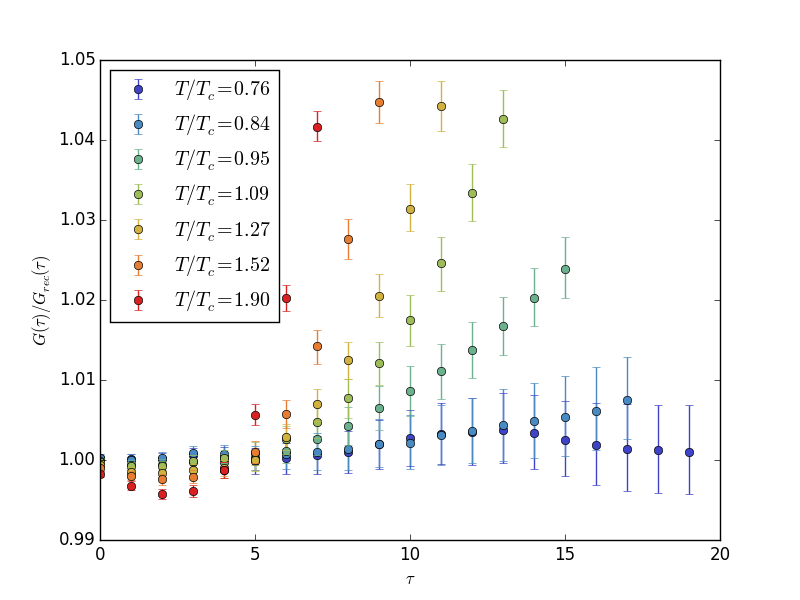}
\caption{The ratio of the charmonium correlator $G(\tau)$ to the reconstructed correlator $G_{rec}(\tau)$ \eqref{Grec-sum} constructed from the $T=T_r=0.24T_c$. Left: pseudoscalar $({\eta}_c)$; right: vector $(J/\psi)$}
\label{fig:GbyGrec-cc}
\end{figure}

\begin{figure} 
\includegraphics*[width=0.5\textwidth]{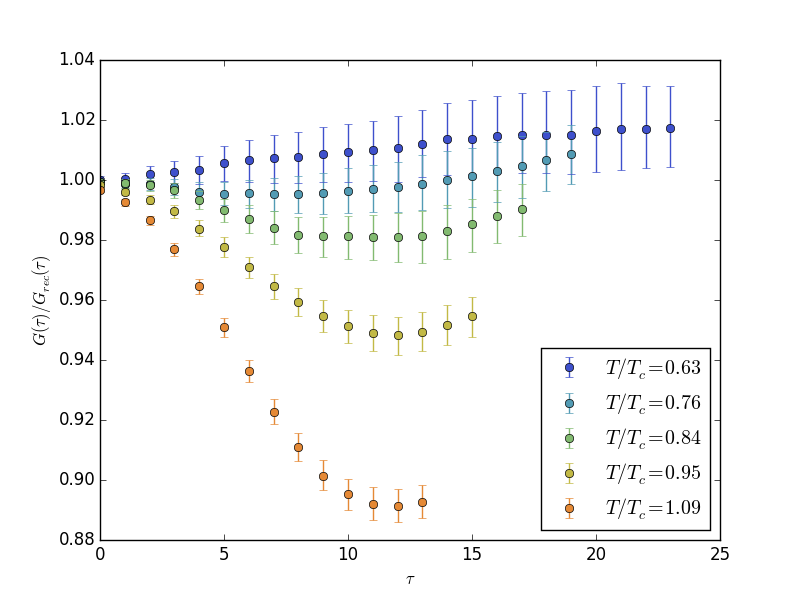}
\includegraphics*[width=0.5\textwidth]{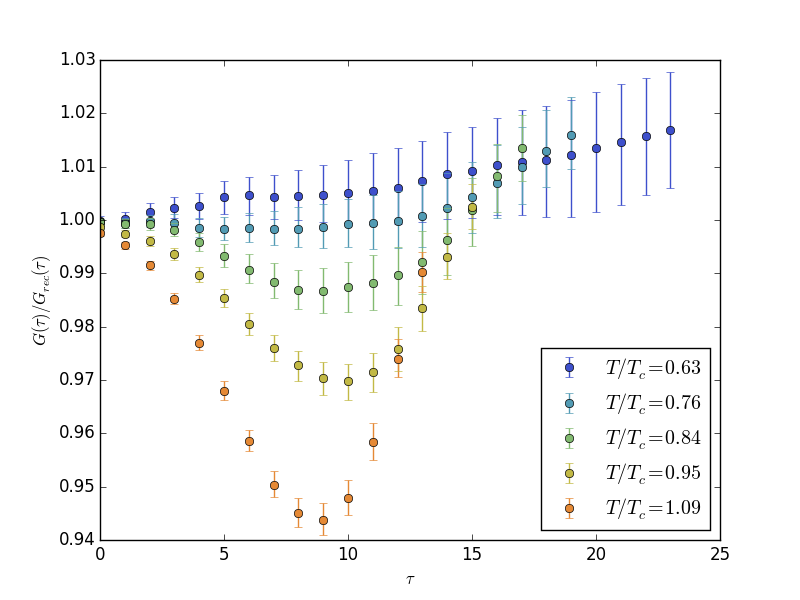}
\caption{As Fig.~\ref{fig:GbyGrec-cc}, for the open-charm pseudoscalar ($D$ meson) channel (left) and vector ($D^*$) channel (right)}
\label{fig:GbyGrec-lc}
\end{figure}

\begin{figure} 
\includegraphics*[width=0.5\textwidth]{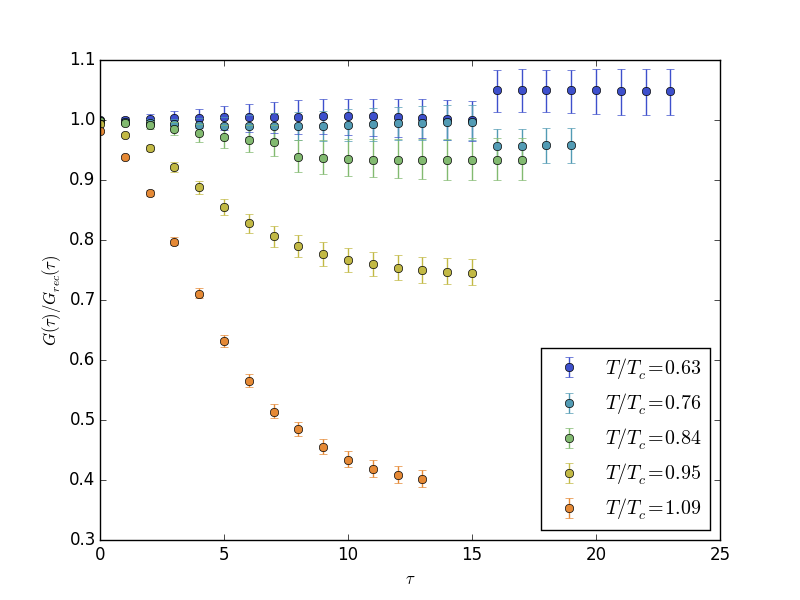}
\includegraphics*[width=0.5\textwidth]{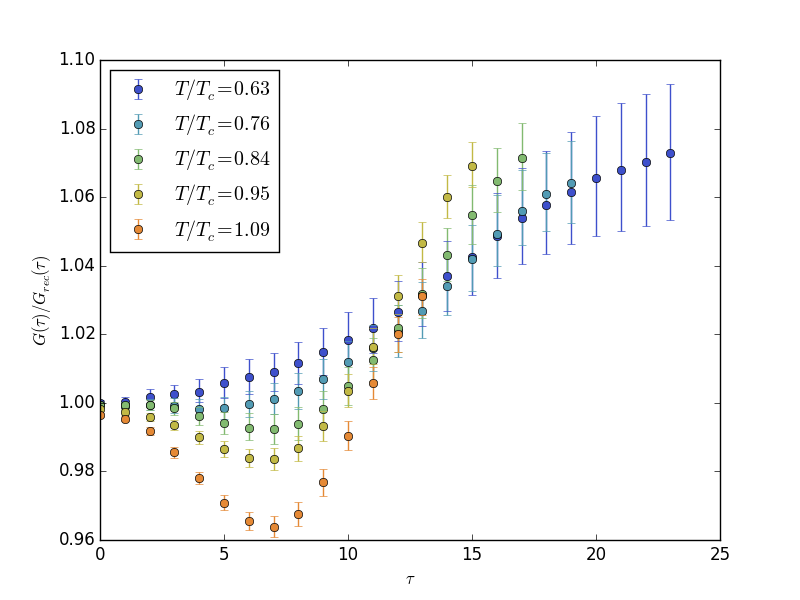}
\caption{As Fig.~\ref{fig:GbyGrec-cc}, for the pion (left) and $\rho$ meson (right)}
\label{fig:GbyGrec-ll}
\end{figure}

We have computed correlators and spectral functions in the vector and pseudoscalar channels for charmonium and open-charm mesons.
In figures \ref{fig:GbyGrec-cc}--\ref{fig:GbyGrec-ll}  we show the ratio of the in-medium correlators to the corresponding reconstructed correlators. This ratio should be 1 if there are no thermal modifications. In the case of charmonium and open-charm we begin to notice thermal modifications at $T=0.84T_{c}$ which become significant above $T=T_c$, in both vector and pseudoscalar channels. For light mesons the changes are already significant at $T=0.84T_{c}$ in the pseudoscalar channel. In all the  vector channels there is an increase in the ratio for high $\tau$. The $J/\psi$ correlator ratio shows a very similar behaviour to that obtained in NRQCD \cite{Aarts:2014cda}, undershooting at small $\tau$ with an upward bend for large $\tau$.

In the pion channel, shown in Fig. 3 (left), we see drastic changes for $T>0.95T_c$, which can be understood as a reflection of the restoration of chiral symmetry, as the pion is no longer a pseudo-Goldstone boson. We see some some jumps in the ratio at low temperatures; these are due to the naive padding method with zeros. We are investigating a more sophisticated method which may remove these errors based on extrapolating the reference correlator to a larger number of points instead of the naive padding with zeros. The $\rho$ meson, shown in Fig.3 (right), has much smaller modifications, comparable in magnitude and shape to those of the $D^*$ meson.

Spectral functions obtained using the BR method are shown in figures 4-7. We have used a constant default model throughout; investigation of the default model dependence of the results is left to a future study. We compare the spectral functions of the correlators with those of the reconstructed correlators to distinguish in-medium effects.  In the case of charmonium this comparison shows that the shifting and broadening of the peak is consistent with the vacuum spectrum up to $1.5T_c$ in the pseudoscalar channel, and $1.9T_c$ in the vector channel. For open-charm we see deviation from the vacuum spectrum as low as $0.76T_c$ in the pseudoscalar channel and $0.84T_c$ in the vector channel, and we see no signs of survival of open charm above $T_c$ in either channel. 

\begin{figure} 
\includegraphics*[width=0.5\textwidth]{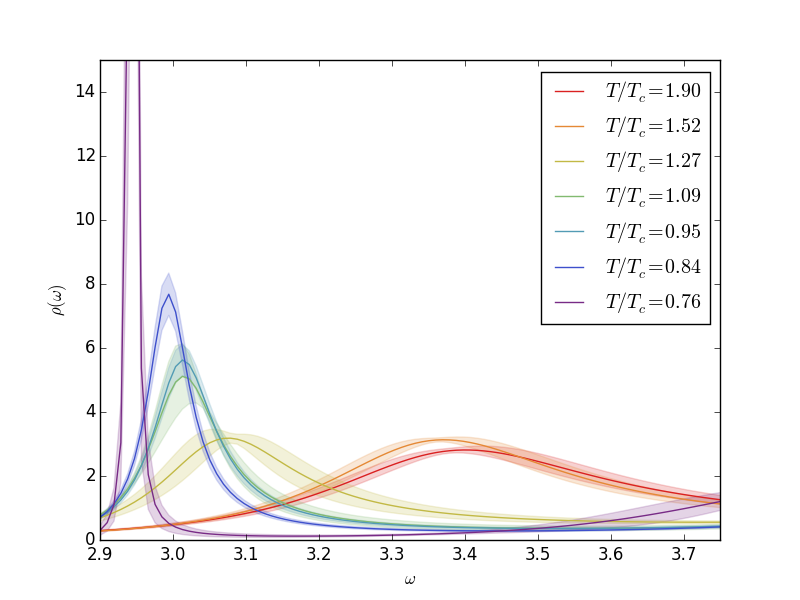}
\includegraphics*[width=0.5\textwidth]{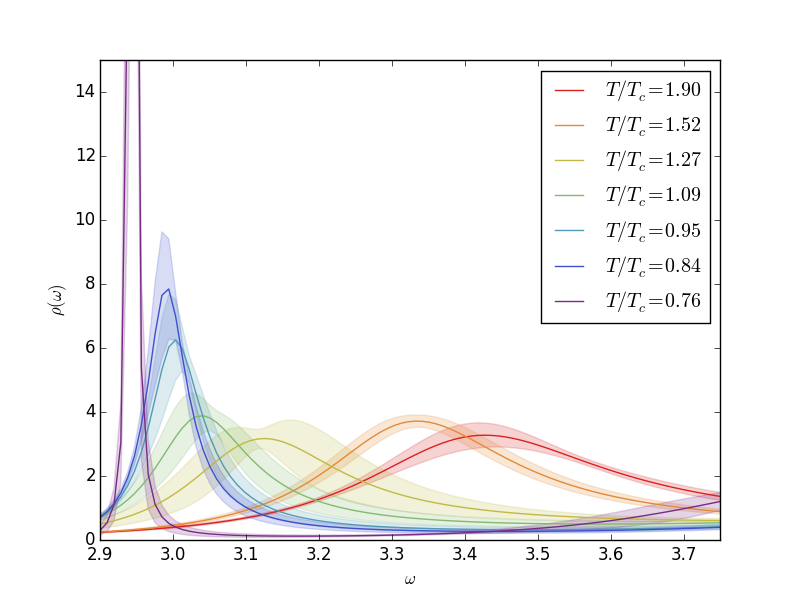}
\caption{Charmonium Spectral function pseudoscalar (correlator - left) (reconstructed correlator - right)}
\end{figure}
\begin{figure} 
\includegraphics*[width=0.5\textwidth]{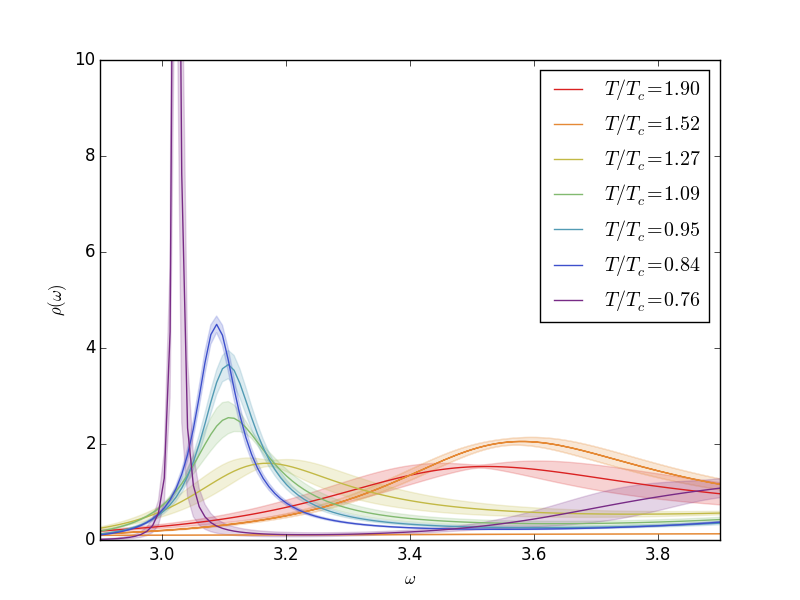}
\includegraphics*[width=0.5\textwidth]{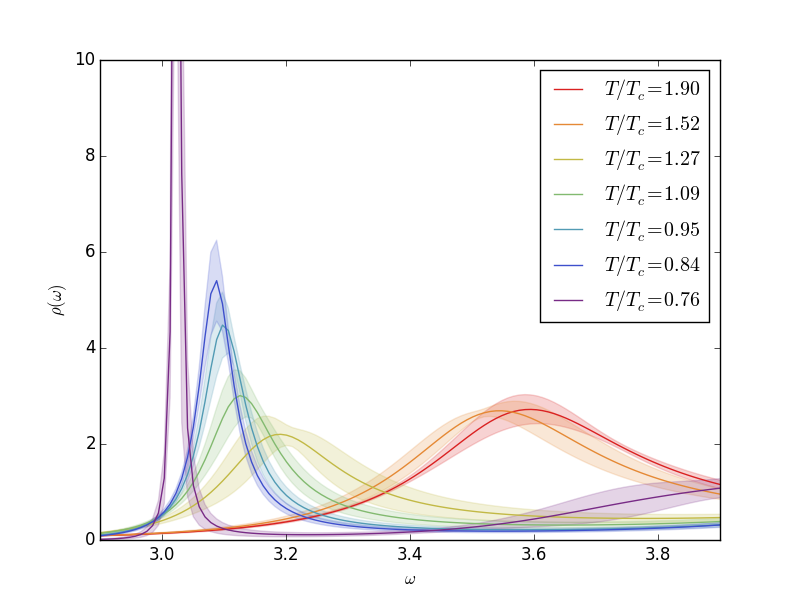}
\caption{Charmonium Spectral function vector (correlator - left) (reconstructed correlator -right)}
\end{figure}
\begin{figure} 
\includegraphics*[width=0.5\textwidth]{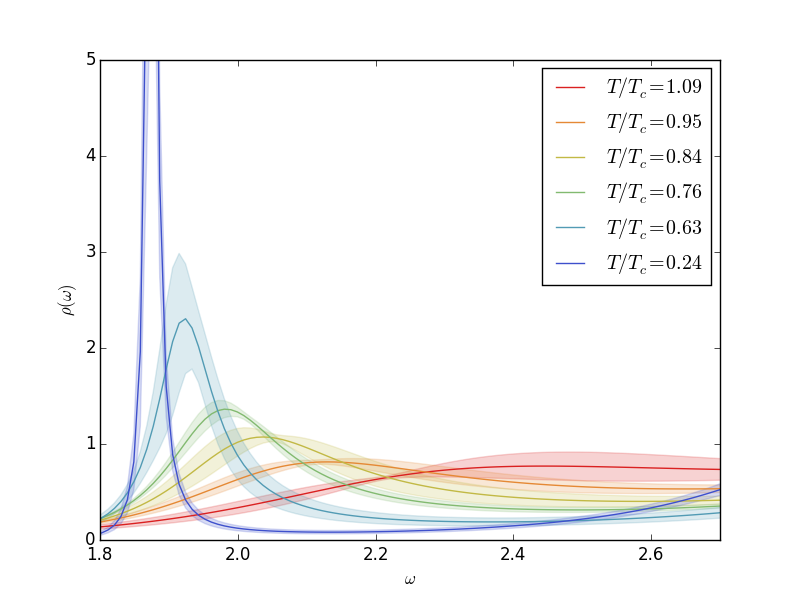}
\includegraphics*[width=0.5\textwidth]{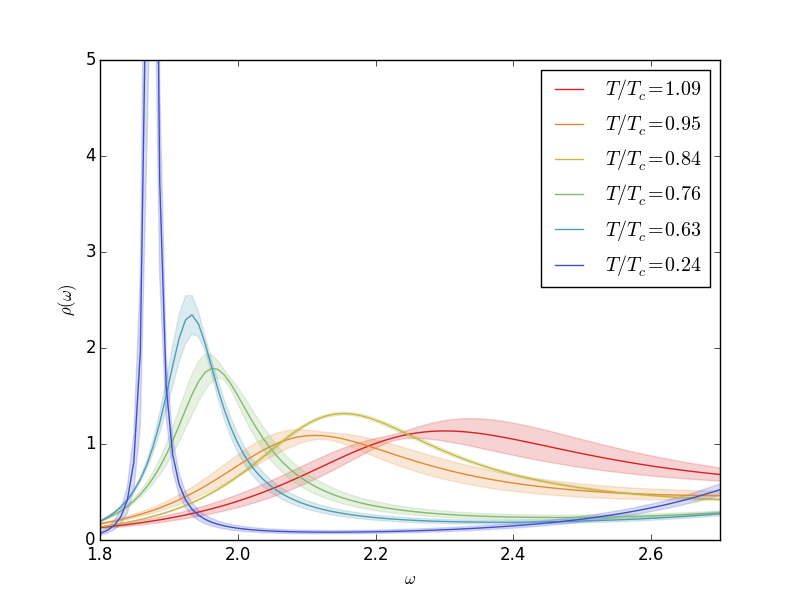}
\caption{Open Charm Spectral function pseudoscalar (correlator - left) (reconstructed correlator -right) }
\end{figure}
\begin{figure} 
\includegraphics*[width=0.5\textwidth]{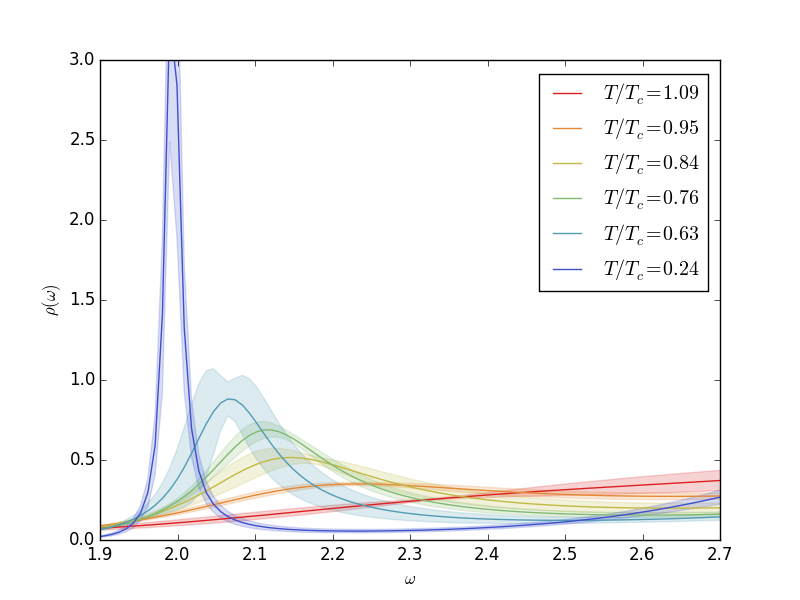}
\includegraphics*[width=0.5\textwidth]{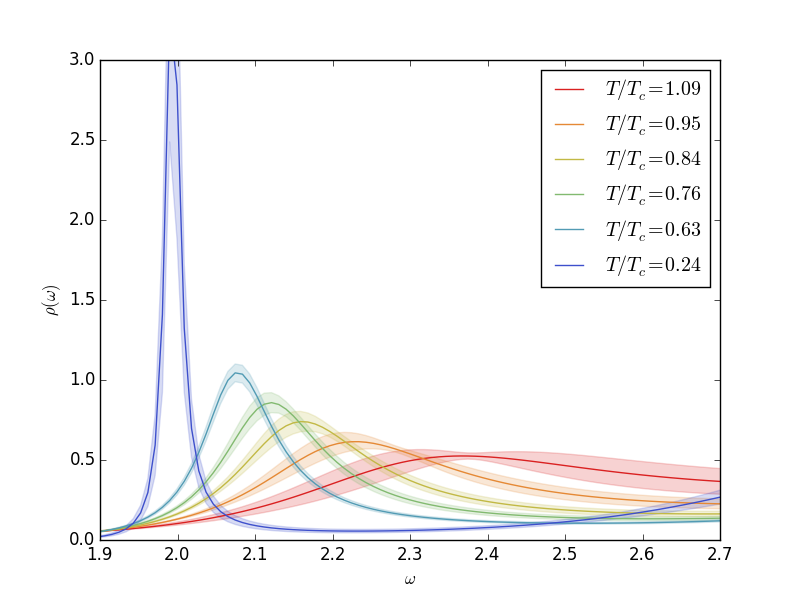}
\caption{Open Charm Spectral function vector (correlator - left) (reconstructed correlator -right)}
\end{figure}

\section{Summary and outlook}

We have studied the temporal correlators and spectral functions of
light, open-charm and charmonium mesons in the pseudoscalar and vector
channels using anisotropic lattice QCD.  We use reconstructed
correlators obtained by direct summation of the zero-temperature
correlator as benchmarks for the thermal behaviour, both by directly
comparing the thermal and reconstructed correlators, and by comparing
the spectral functions obtained from the termal correlators to those
obtained from the corresponding reconstructed correlators.

We see significant changes in the pion correlator as we approach the
pseudocritical temperature $T_c$, consistent with the restoration of
chiral symmetry.  In the open-charm channels we find significant
thermal modifications in the correlators already below $T_c$, and no
sign in the spectral functions that the open-charm mesons may survive
in the quark-gluon plasma.

In the charmonium ($J/\psi$ and $\eta_c$) channels, we find small but
significant modifications in the correlators already from $T\sim
T_c$.  However, the spectral functions remain consistent with the
vacuum (reconstructed correlator) spectral functions up to $1.5T_c$ in
the vector channel and $1.9T_c$ in the pseudoscalar channel.

It is worth noting that in all channels, we see that the ground state
peak broadens and shifts to higher frequencies as the temperature
increases.  However, this behaviour is found for both thermal and
reconstructed correlators, and must therefore be interpreted as an
effect of the limited number of temporal points available, rather than
a physical effect.

In the future, we plan to extend this study to charmonium P-waves and
also attempt to determine transport properties (conductivity and
diffusion) from the high-temperature spectral function.  The results
shown here have all been obtained using the BR method; a detailed
comparison of MEM and the BR method as well as other methods,
including a systematic study of model function dependence, is work
in progress.

\section*{Acknowledgments}
This work has been carried out using computational resources provided
by the Irish Centre for High End Computing and the STFC funded DiRAC
facility.  We acknowledge the networking support by the COST action
CA15213 ``Theory of hot matter and relativistic heavy-ion
collisions''.  RQ has been supported by a Maynooth University SPUR
scholarship.

\bibliography{ref}
\end{document}